\documentclass[referee]{aa} 

\usepackage{graphicx}
\usepackage{tabularx}
\usepackage{amsmath}
\usepackage[colorlinks=true,linkcolor=magenta,citecolor=blue,filecolor=cyan,urlcolor=purple,final=true]{hyperref}
\usepackage{txfonts}
\usepackage{ulem}

\begin{document}

    \title{The Automatic Identification and Tracking of Coronal Flux Ropes - Part II: New Mathematical Morphology-based Flux Rope Extraction Method and Deflection Analysis}

   \author{A. Wagner
          \inst{1,2},
          S. Bourgeois
          \inst{3,4}
          \and
          E. K. J. Kilpua
          \inst{1}
          \and
          R. Sarkar
          \inst{1}
          \and
          D. J. Price
          \inst{1}
          \and
          A. Kumari
          \inst{5}
          \and
          J. Pomoell
          \inst{1}
           \and
          S. Poedts
          \inst{2,6}
            \and
          T. Barata
          \inst{7}
           \and
          R. Erd\'elyi
          \inst{4,9,10}
           \and
          O. Oliveira
          \inst{8}
          \and
          R. Gafeira
          \inst{3}}

    \institute{Department of Physics, University of Helsinki, P.O. Box 64, FI-00014, Helsinki, Finland\\
    \email{andreas.wagner@helsinki.fi}
    \and
    CmPA/Department of Mathematics, KU Leuven, Celestijnenlaan 200B, 3001 Leuven, Belgium
    \and
    Instituto de Astrof\'isica e Ci\^{e}ncias do Espa\c{c}o, Department of Physics, University of Coimbra, Coimbra, Portugal
    \and
    Solar Physics and Space Plasma Research Centre (SP2RC), School of Mathematics and Statistics, University of Sheffield, Sheffield, S3 7RH, United Kingdom
    \and
    NASA Goddard Space Flight Center, Greenbelt, MD 20771, USA
    \and
    Institute of Physics, University of Maria Curie-Sk{\l}odowska, ul.\ Radziszewskiego 10, 20-031 Lublin, Poland 
    \and
    Instituto de Astrof\'isica e Ci\^{e}ncias do Espa\c{c}o, Department of Earth Sciences, University of Coimbra, Coimbra, Portugal
    \and
    CFisUC, Department of Physics, University of Coimbra,  3004 516 Coimbra, Portugal
    \and
    Department of Astronomy, E\"otv\"os Lor\'and University, P\'azm\'any P\'eter s\'et\'any 1/A, H-1117 Budapest, Hungary 
    \and
    Gyula Bay Zoltan Solar Observatory (GSO), Hungarian Solar Physics Foundation (HSPF), Pet\H{o}fi t\'er 3., H-5700 Gyula, Hungary  
    \\
    }
   \date{Received XXX; accepted XXX}

\abstract
{Constructing the relevant magnetic field lines from active region modelling data is crucial to understand the underlying instability mechanisms that trigger the corresponding eruptions.}
{We present a magnetic flux rope (FR) extraction tool for solar coronal magnetic field modelling data, which builds upon the methodology from \cite{Wagner2023}. 
The newly developed method is then compared against its previous iteration. Furthermore, we apply the scheme to magnetic field simulations of active regions AR12473 (similar to our previous study) and AR11176. We compare the method to its predecessor and study the 3D movement of the newly extracted FRs up to heights of 200 and 300 Mm, respectively.}
{The extraction method is based on the twist parameter $T_w$ and a variety of mathematical morphology (MM) algorithms, including the opening transform and the morphological gradient. We highlight the differences between the methods by investigating the circularity of the FRs in the plane we extract from.
The simulations for the active regions are carried out with a time-dependent data-driven magnetofrictional model (TMFM; \cite{Pomoell19}). We investigate the FR trajectories by tracking their apex throughout the full simulation time span.}
{Comparing the newly developed method to the extraction scheme in \cite{Wagner2023}, we demonstrate that this upgrade provides the user with more tools and less a-priori assumptions about the FR shape that, in turn, leads to a more accurate set of field lines. Despite some differences, both the newly extracted FR of AR12473 as well as the FR derived with the old iteration of the method show a similar general appearance, confirming that both methods indeed extract the same structure. The methods differ the most in their emergence/formation stages, where the newly extracted FR deviates significantly from a perfectly circular cross-section (that was the basic assumption of the initial method). The propagation analysis yields that the erupting FR from AR12473 showcases indeed stronger dynamics than the AR11176 FR and a significant deflection during its ascent through the domain. The modelling results are also verified with observations, with AR12473 being indeed dynamic and eruptive, while AR11176 only features an eruption outside of our simulation time window.}
{We implemented a FR extraction method, incorporating mathematical morphology algorithms for 3D solar magnetic field simulations of active region FRs. This scheme was applied to AR12473 and AR11176. We find that the clearly eruptive FR of AR12473 experiences significant deflection during its rise. The AR11176 FR appears more stable, though there still is a notable deflection. This confirms that at these low coronal heights, FRs do undergo significant changes in the direction of their propagation even for less dynamic cases.}

   \authorrunning{A.~Wagner, et al.}
   \titlerunning{The Automatic Identification and Tracking of Coronal FRs II}
   \maketitle

\section{Introduction}
\label{Sect: Intro}
Solar eruptions play a central role in many space weather studies, as their space weather effects have direct impact on our technology and life on Earth. In particular, coronal mass ejections \citep[CMEs; e.g.][]{Webb2012} can cause great disruptions to both our space- and ground-based technology, and related high energy particle radiation can pose a risk to astronauts and high-altitude airplane crew and passengers \citep[see e.g.,][]{Zhang2021, Bain2023}. Thus, there is broad interest in the understanding of the triggering mechanisms of solar eruptive phenomena and their evolution from initiation until their arrival at Earth.

Flux ropes (FRs) are the key magnetic structures related to CME eruptions, \citep[see e.g.,][]{Vourlidas2013,Green2018,Chen2017}. Due to the limited capability of observing the magnetic field in the solar corona, modelling the corona via numerical simulations is necessary to study the formation, destabilization and dynamics of solar FRs. There are a wide range of numerical simulations that incorporate physics to varying degrees in many different implementations, like potential field methods, non-linear force-free fields or magnetohydrodynamic (MHD) models, to name a few \citep[e.g.,][and references therein]{Kilpua2019}. Identifying and tracking of the magnetic FR (i.e., following the field lines that belong to the same magnetic structure) that are either injected into these simulations or directly result from them is not a trivial task. 

Often, very approximate methods are used to determine the magnetic field lines of FRs from the models. Manual extractions are widely performed \citep[e.g.,][]{Kumari2023} and only few automatic methods exist, for example \cite{Lowder17}, who have constructed a FR identification and tracking procedure for coronal magnetic field simulations. The authors base their method on thresholding photospheric maps of field line helicity, while the tracking is carried out by calculating the overlap of footpoints between two output frames. 

In this paper, we present a novel, semi-automatic FR extraction scheme to identify and track the evolution of solar FRs in coronal simulations. Here, we base the extraction on the twist parameter $T_w$ \citep[e.g.,][]{Berger2006, Liu16}, paired with a nonlinear image processing method called mathematical morphology (MM). A prominent advantage of the MM methodology is that any proxy that captures the relevant FR field lines can be employed. The incorporation of MM methods was chosen because it allows the user to have more control over the FR extraction procedure, as well as to remove the need for assumptions to simplify the process. This new extraction scheme builds upon the methodology developed in \cite{Wagner2023}, to which we will be referring to as "Paper~I" hereafter. The previous method was also focused around extracting a FR from the $T_w$ maps. Instead of MM algorithms aiding the procedure, the assumption of a perfectly circular cross-section in the plane of extraction was employed. Additionally, two parameters were introduced to control the FR size depending on the gradient in the twist maps ($\kappa$) and the amount of overlying, non-FR twisted field lines ($\epsilon$). The tool presented in this paper removes both the restricting circularity assumption as well as complications due to the additional parameters. The new tools will be tested on the output of a time-dependent data-driven magnetofrictional model \citep[TMFM;][]{Pomoell19} applied to active regions (ARs) AR12473 (passed the center of the solar disk on 28 Dec 2015) and AR11176 (passed the center of the solar disk on 28 Mar 2011). The former AR is the same as analysed in Paper~I, which allows us to directly compare the results and highlight the improvements of the new scheme, while AR11176 serves as an additional case to test the methodology. 

The mathematical morphology method was introduced by \cite{matheron1967} and \cite{serra1969} who applied it to probe porous media \citep{Serra1982,Matheron2001}. Based on set theory, topology, and integral geometry, MM has turned out to be a powerful tool to describe the geometrical shape of objects of interest in images. According to \cite{Carvalho2020}, MM is particularly suited for analyzing structures with complex and irregular shapes and/or sizes. Quantitative properties of these structures can be easily found with specific implementations provided by MM software libraries. Moreover, MM is applicable to different types of images, from low to higher resolution as it can deal with images taken at both ground-based as well as space-borne observatories. Therefore, it is also a suitable methodology for the field of heliophysics. Several studies have already demonstrated that the MM analysis is able to characterise a vast range of solar features accurately. For example, \cite{Shih2003}, \cite{Qu2005} and, more recently, \cite{Koch2015} applied MM techniques to detect solar filaments in H-alpha full-disk and far-infrared solar images, while \cite{Teresa2018} and \cite{Carvalho2020} uncovered sunspots and solar plages through a MM algorithm pipeline. \cite{Stenning2013} used MM to classify sunspots. 

In this paper we apply MM operations to extract and process the early-phase FR structures from the simulation generated twist maps.
In order to obtain the external outlines of these structures, we implement an operator called the \textit{morphological gradient} (cf., Sect.~\ref{Sect: MM}). 

The paper is organized as follows: In Sect.~\ref{Sect: Data} we discuss the data and methods, needed to perform the simulation and extraction. Sect.~\ref{Sect: Results} showcases the resulting flux ropes and their trajectories. Finally, Sect.~\ref{Sect: Discussion} discusses the method's performance and the results of the extraction and deflection analysis. 

\section{Data and Methods}
\label{Sect: Data}
\subsection{Observational Data}
The TMFM simulation requires a time series of photospheric vector magnetograms (see Sect.~\ref{Sect: TMFM} below). We therefore use magnetic field data from the Helioseismic and Magnetic Imager \citep[HMI, see][]{Couvidat16,Schou12} from the Solar Dynamics Observatory (SDO) \citep[see][]{Pesnell12} mission. 

To compare with our modelling efforts, we furthermore use image material from the Atmospheric Imaging Assembly \citep[AIA, see][]{Lemen12} instrument of SDO. In particular, we are making use of the 171, 193 and 211~\AA{} channels.  

\subsection{The TMFM Model and Set-up}
\label{Sect: TMFM}
Analogously to Paper~I, we apply a time-dependent data-driven magnetofrictional model \citep[TMFM;][]{Pomoell19} to active region AR12473. In Paper~I, we paired this simulation with a so-called relaxation procedure, where the TMFM was coupled to a zero-beta magnetohydrodynamics approach \citep{Daei2023}. In addition, in this work we also apply the tracking methodology to a second active region, AR11176. 
From the HMI vector magnetograms, electric field maps are created, which are then ultimately used as input to the model (more details on this procedure can be found in \cite{Lumme17} and \cite{Pomoell19}). 
The dynamics in the model is set based on the assumption of the velocity $\mathbf{v}$ in the coronal domain being proportional to the Lorentz force:
\begin{equation}
   \mathbf{v} = \frac{1}{\nu} \frac{\mu_0 \mathbf{J} \times \mathbf{B}}{B^{2}},
\end{equation}
with $\mathbf{B}$ the magnetic field, $\mu_0$ the vacuum magnetic permeability, $\mathbf{J}$ the current density and $\nu$ the magnetofrictional coefficient. 

The TMFM simulation of AR12473 is performed for the time interval from 22 Dec 2015, 23:36 UT to 2 Jan 2016, 12:36 UT, thus identical to Paper~I, and of AR11176 from 25 Mar 2011 at 4:00 UT until 1 Apr 2011, 18:00 UT. These periods correspond to the times when the active regions were best observed on-disk and not too close to the limbs of the Sun, which is necessary for keeping projection effects in the input data minimal. The simulations are both analysed at a cadence of 6 hours. The Cartesian domain for AR12473 has an approximate size of 386x254x200~Mm while the domain for AR11176 has approximate dimensions of 800x374x300~Mm (corresponding to $x$, $y$ and $z$, respectively, with the plane of the magnetogram being the $z=0$ plane). In total, the AR12473 simulation domain consists of approximately 10.6 million cells, while the AR11176 domain contains approximately 48.8 million cells. For AR11176 the domain height is larger (300~Mm compared to 200~Mm)  due to specifics of the FR that will be detailed later in Sect.~\ref{Sect: AR11176}. One fundamental difference between the two sets of simulations is the masking of the magnetograms: while for AR12473 no masking is applied (analogously to \cite{Price20} and \cite{Kumari2023}), for AR11176 we set all $B$-field values below 250~G to zero. This step had to be done for AR11176, as the simulation with an unmasked magnetogram yielded only a small, stable and weakly twisted bundle of magnetic field lines close to the photosphere. Otherwise, the pre-processing steps are analogous to \cite{Price20}. 

\subsection{Mathematical Morphology (MM) Algorithms}
\label{Sect: MM}
We use MM algorithms (introduced in Sect.~\ref{Sect: Intro}) to aid our extraction method in multiple ways. MM is built upon a reference object called the \textit{Structuring Element} (SE). This object has a size, shape, and orientation which are a-priori defined by the user, and it is used as a kernel to probe images. Most MM transforms operate by comparing the structures of interest in images to this carefully user-selected SE. In other words, we extract, reject or modify image structures that the SE fits or misses, respectively - hence the importance of accurately choosing an appropriate shape and size for the SE \citep{Soille1999}. 

\textit{Erosion} and \textit{dilation} are the two fundamental MM operations that provide a basis for all other image transforms. On the one hand, the erosion transform is a useful tool in morphological image processing to enhance the darkest regions of an image, called \textit{valleys}, and to numb the intensity of the brightest ones, called \textit{peaks} \citep{Teresa2018}. From the point of view of set theory, the erosion of a set $X$, $\epsilon_{A}(X) = X \ominus A$, is determined by all the centre points $x$ of the SE $A$, such that $A$ is included in $X$. Thus, for binary images, this has the effect that pixels around the object boundaries may be removed. On the other hand, the dilation transform of a set $X$, $\delta_{A}(X) = X \oplus A$, is defined as the union of all the centre points $x$ of the SE $A$, where $A$ and $X$ have non-zero overlap. Contrary to erosion, the dilation operator adds pixels around the original object boundaries and widens the bright areas while reducing the valleys \citep{Teresa2018}.

The morphological gradient subtracts the erosion from the dilation for a grey-scale image $f$: $\nabla(f) = \delta_{A}(f) - \epsilon_{A}(f)$, also written as $ \nabla(f) = (f \oplus A) - (f \ominus A)$ where $f$ is the image and $A$ the SE. The morphological gradient helps to accentuate variations in the twist number maps, as will be shown in Sects.~\ref{Sect: Extraction} and \ref{Sect: Discussion}.

\subsection{Extraction Method based on MM}
\label{Sect: Extraction}
Following the procedure employed in Paper~I, the extraction starts from a 2D slice, chosen to cut the domain close to the polarity inversion line (PIL) of the simulated active region. In theory, any plane where the FR is expected to pass through, is suitable. In this slice, we calculate the twist map using the twist number $T_w$, according to the approach in \cite{Liu16}. $T_w$ counts the number of turns two infinitesimally close field lines make about each other \cite{Berger2006}. More discussion on the choice of $T_w$ as the FR proxy can be found in Sect.~\ref{Sect: Tw}. Figure~\ref{fig: Twist AR12473} shows as an example $T_w$ maps, reconstructed for a few selected frames from the TMFM simulation output for AR12473 (the plane is placed at $x \approx 0.7$~Mm). These maps have a resolution of approximately 0.36 Mm per pixel for both active regions.

In the next step, we apply the morphological gradient to sharpen the twist maps. This has the effect that the subsequent application of a threshold is more stable with regards to the resulting outlines. This allows us to use a constant threshold throughout the whole time series to identify high-twist regions (and ultimately reduce the twist maps to binary masks) without removing relevant parts or creating unwanted artefacts. Given our twistmap resolution, we used circular SEs, with sizes between 4 and 10 for this procedure.
In the resulting binary masks, we then seek to improve our extraction on a case-by-case basis. Indeed, it may be necessary to smooth out the extracted structures and/or to remove any spur or twisted feature that would not be part of the actual FR. To do so, we apply a morphological \textit{opening}. The opening transform of an image $f$, $\gamma_{A}(f)$, is defined as the dilation of the erosion of the image $f$, i.e., $\gamma_{A}(f) = \delta_{A}(\epsilon_{A}(f))$, also written as $ \gamma = f \circ A = (f \ominus A) \oplus A$. 
Generally, the initial structure is not entirely recovered since erosion and dilation are not reversible. Thus, the opening is effective in our case for filtering noise out of the initial FR extraction, depending on the choice for the structuring element. One must be very careful when applying an opening with a large SE, as it may irreversibly suppress details that are yet relevant for an accurate extraction. To prevent this, we only perform opening operations on the frames that require them and, in particular, only to sub-regions of these frames where noise and unwanted structures are visible. After trial and error, we find the optimal SE for each time frame by incrementally increasing the SE size and checking the resulting extraction by visual inspection. This approach ensures that we do not use unnecessarily large SE sizes. We furthermore exclusively use circles for the shape of our SE. The SE sizes range from 10 to 50 for early-to-mid stages of the AR12473 simulation, while we had to use substantially larger SE sizes ($> 100$) for the late-stages, as there was a significant connection with a non-FR feature in the maps. For the AR11176 simulation, on the other hand, SE sizes of approximately 10 were sufficient.
Once the outlines appear satisfactory, we proceed with the tracking, which is implemented similarly to Paper~I, that is, the region of interest is identified as the same structure in subsequent frames if they overlap sufficiently. 

Our method also offers some post-processing routines, for example the erosion algorithm. Drawing the analogy to Paper~I: the size of the SE of the erosion here is analogous to the $\epsilon$-prescription, that is, a parameter determining to which degree the extracted shape will be eroded. The difference is that the morphological erosion works for any arbitrary shape, while our previous $\epsilon$-method per definition was only applicable for circular shapes. One may also use a dilation operation as a post-processing step to fill holes in the extracted area. Finally, we compute the source points from the resulting outlines as uniformly distributed points within the derived binary FR mask. We note that, in practice, if no frame-specific pre- or post-processing is needed, the extraction is practically automatic, once the relevant parameters have been set. 

To qualitatively showcase the result of the MM-based method in contrast to the result of the method presented in Paper~I, we calculate, for both AR11176 and AR12473, the circularity of the FR cross-section. It is the ratio of the standard deviation of the FR radius (defined as the distance between the centroid of the extracted shape to its edge pixels) over its mean radius. In other words, we seek to estimate the variations of the FR radius throughout the whole structure, with a circularity of zero indicating that the FR shape is a perfect circle. For the original implementation to work best, a circularity close to zero is required, which we will probe with our new extraction results. We furthermore calculate the circularity for an additional FR extraction, where we do not apply any post-processing (except for one problematic frame in each simulation to ensure the successful tracking of the FR), to show the effect of these algorithms on the circularity analysis.

\subsection{FR Propagation and Deflection}
\label{Sect: Deflection}
Having full access to the FR field lines enables, for example, detailed studies of the FR trajectory through the simulation domain. We investigate here the FR trajectory in terms of the angle of propagation (and with that the deflection) within the simulation domain as a function of time, as well as the evolution of all spatial coordinates. To achieve this, we calculate the FR apex as follows: First, we compute the maximum height that any FR field line, for a given frame, reaches. Then, we define the FR apex as the average over all field lines that reach within 15~Mm of the maximum height. This procedure mitigates abrupt jumps in the horizontal components the FR apex may exhibit if many high reaching field lines are present. The results are given here as a function of the simulation frame number with the time difference between each frame being six hours (determined by the time cadence of the output).

We track the apex from the moment the FR forms (i.e., when there is a coherent bundle of field lines with sufficient twist ($T_w \gtrapprox 0.8$ for AR12473 and $T_w \gtrapprox 0.4$ for AR11176) present), until it reaches the top of the simulation domain or before it starts interacting with it. To measure how much the FR trajectory varies during its propagation we define the propagation angle as the angle between the $z$-axis and the vector that extends from the first apex point to the apex point at the time step in question. The larger the propagation angle is, the more the FR propagation direction differs from a trajectory parallel to the $z$-axis (which, in the spherical case, would correspond to the radial direction).

\section{Results}
\label{Sect: Results}
\subsection{Analysis of AR12473 simulation}
\label{Sect: AR12473}
\begin{figure}
    \centering     
    \includegraphics[width=0.8\linewidth]{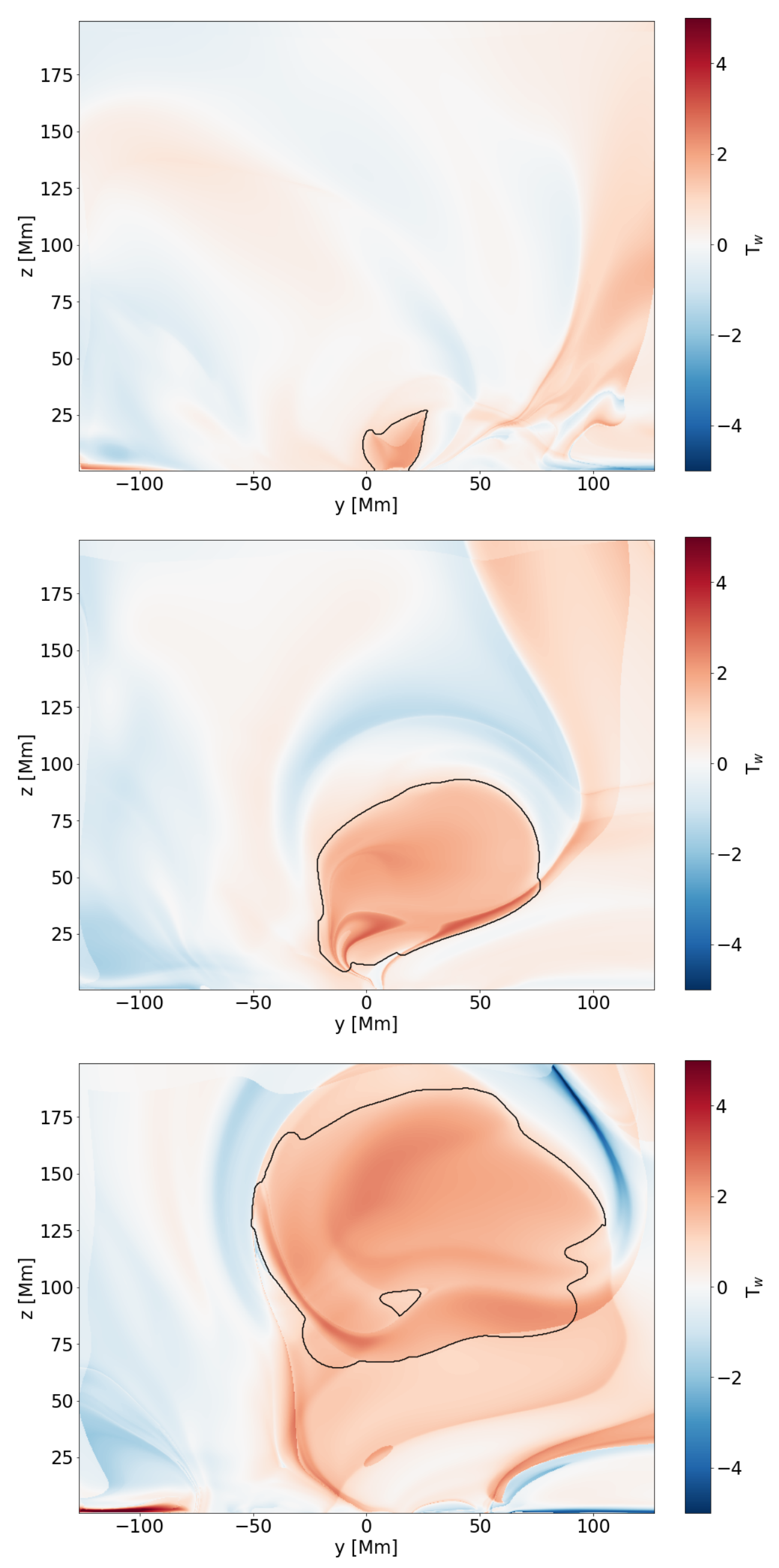}
    \caption{Twist map evolution for AR12473. The colour intensity indicates the magnitude of twist, where blue indicates negative twist and red indicates positive twist. The shown snapshots correspond to simulation frame numbers: 7, 17 and 27 (from top to bottom). The shape, identified as belonging to the FR is marked with a black contour.}
     \label{fig: Twist AR12473}
\end{figure}

A selection of snapshots from the time series of $T_w$-maps are shown in Fig.~\ref{fig: Twist AR12473} (the plane is located at $x \approx 0.7$ Mm). The movie showing the full evolution of the twist map is included as Supplementary material. The twist maps show that the FR forms within the first frames, corresponding to $\approx 1.5\;$days close to the bottom of the domain and then starts to rise upward. The early formation phase is shown in the top panel of Fig.~\ref{fig: Twist AR12473}, when there is only a weakly twisted (faintly reddish) structure at $y = 0$. When the simulation progresses, this structure evolves into a coherent, twisted FR that moves upward. 

\begin{figure*}
     \centering
     \includegraphics[width=\linewidth]{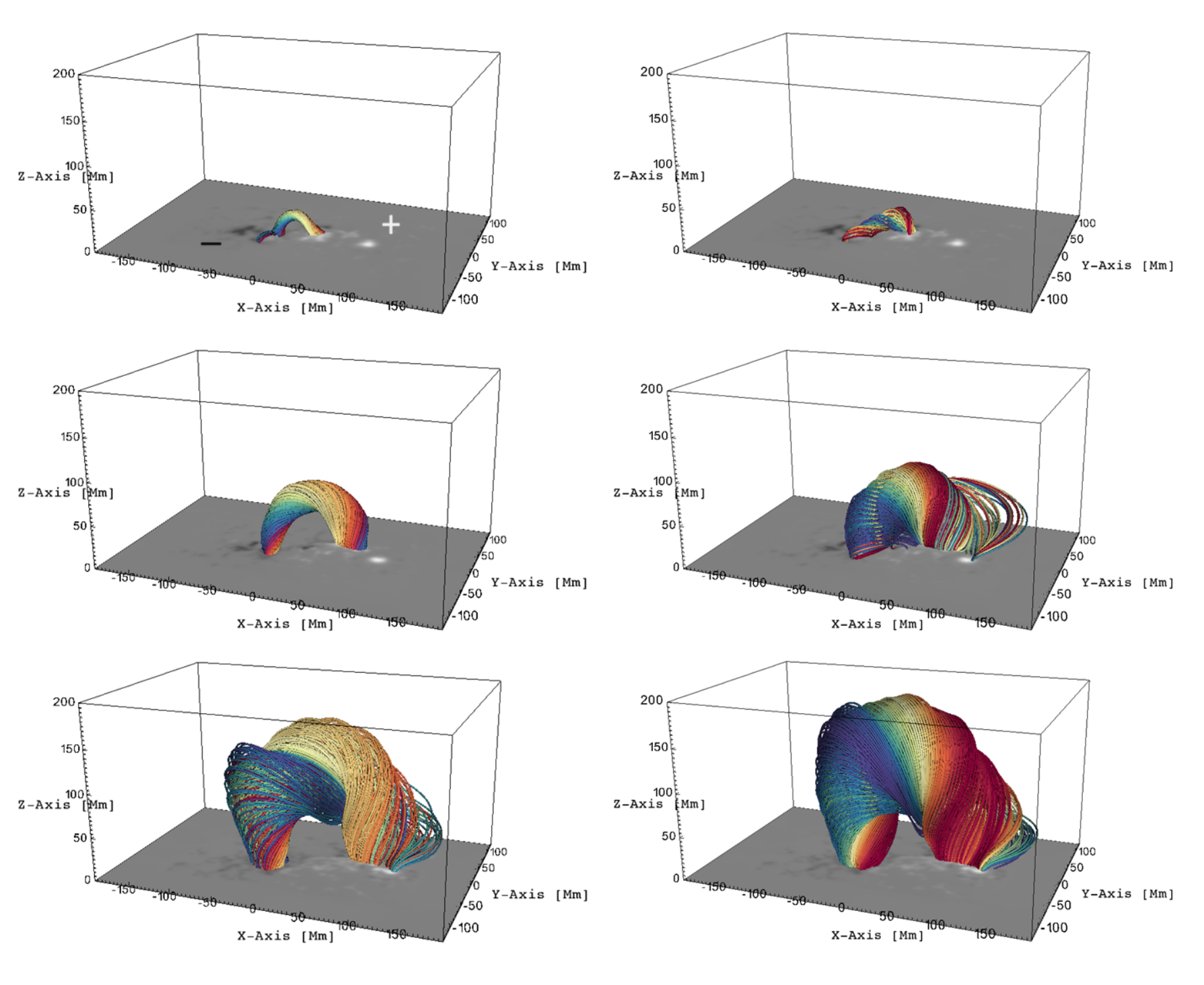}
     \caption{FR snapshots for the TMFM AR12473 simulation, using the method from Paper~I (on the left), versus the presented extraction method (on the right). As in Fig.~\ref{fig: Twist AR12473}, the frames are from top to the bottom: 7, 17, 27. The magnetic polarity of the different footpoint regions is labelled in the top left panel in the corresponding colour (black = negative, white = positive).}
     \label{fig: AR12473}
\end{figure*}

The FR field lines for AR12473 are extracted following the scheme described in Sect.~\ref{Sect: Extraction}, and are shown in Fig.~\ref{fig: AR12473} on the right, compared against the extraction from Paper~I with an $\epsilon$ value of 0.03, using the same simulation data and frames (left panels). Comparing the results of the two methods, one can see that throughout all stages of the evolution of the structure, the FR appears significantly thicker when extracted with the new method. The FRs show clear differences also in their overall coherence and appearance for the earlier frames 7 and 17. For the mid-frame, the key difference between the two extraction scheme outputs is (in addition to thickness) that for the FR extracted using the MM-based method there is a set of field lines included that connects to the adjacent positive polarity region located at approximately $(x, y, z) = (100, 0, 0)$ and visible in the middle panel on the right and indicated by a plus in the first panel (top left). 
In the last frame shown, the field lines both from the old and updated method extend to this region, indicating that this region is indeed tied to the FR. Additionally, the FRs are more similar in this frame than at earlier times. A closer look at the footpoint regions also reveals that the positive polarity footpoint exhibits significant movement throughout the simulation, while the negative polarity footpoints approximately stay rooted at the same location in the photosphere. This confirms also our findings from Paper~I, where this was investigated in great detail. 

\begin{figure*}
     \centering     
     \includegraphics[width=\linewidth]{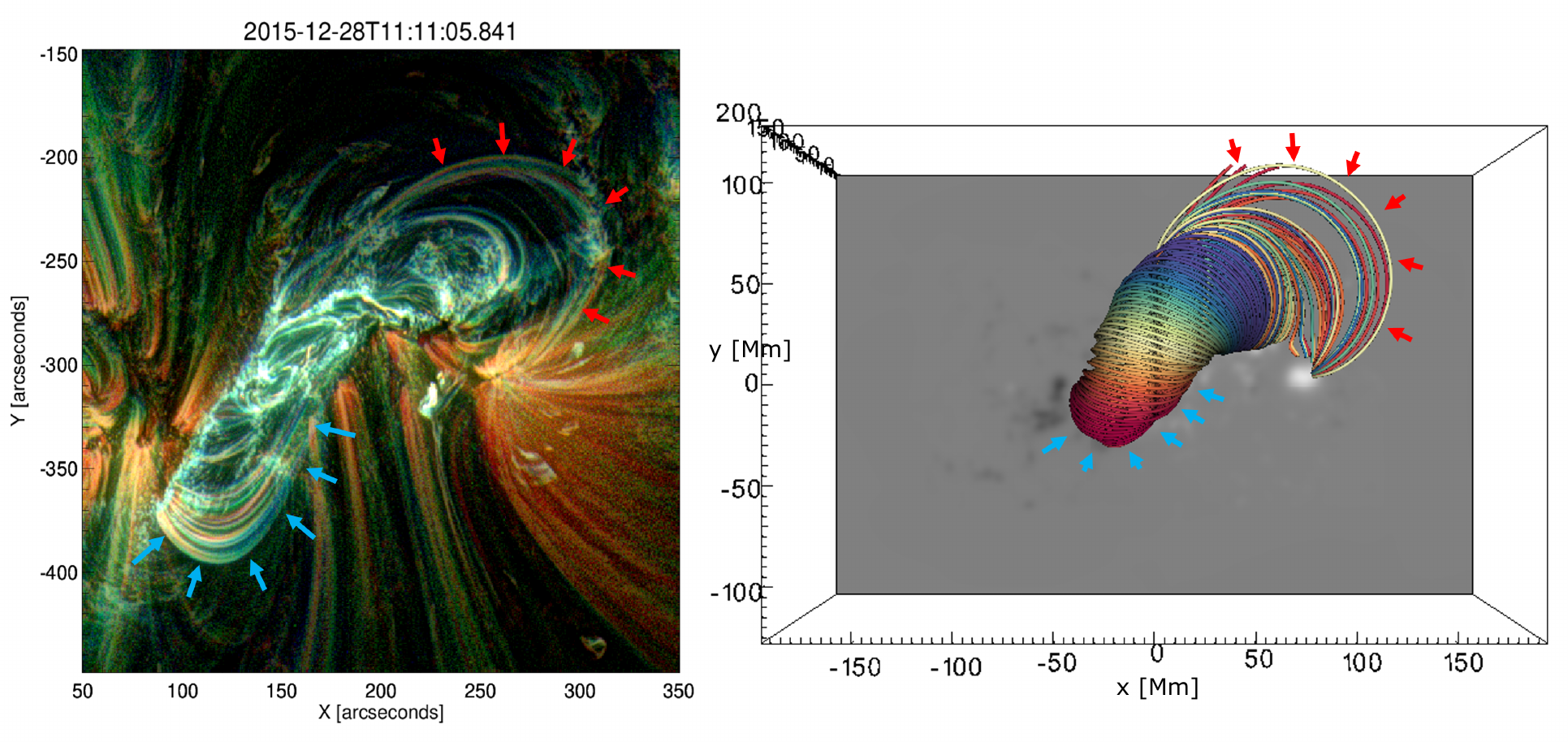}
     \caption{Comparison of FR appearance as observed in EUV (left panel) and the simulation (right panel) for AR12473. The EUV image is the composite image constructed from the AIA 171 \AA\ (red), 211 \AA\ (green) and 193 \AA\ (blue) passbands. Blue and red arrows indicate similar structures between the AIA images and the simulation results.}
     \label{fig: Aia_model_comparison}
\end{figure*}

The above findings highlight the fact that the updated method captures more relevant field lines, while still avoiding surrounding non-FR field lines. A top-view comparison of the FR field lines using the MM extraction to AIA observations can be seen in Fig.~\ref{fig: Aia_model_comparison}, which shows that  the overall morphology of the observed structure is well captured. 

\subsection{Analysis of AR11176 simulation}
\label{Sect: AR11176}

\begin{figure*}
     \centering     
     \includegraphics[width=\linewidth]{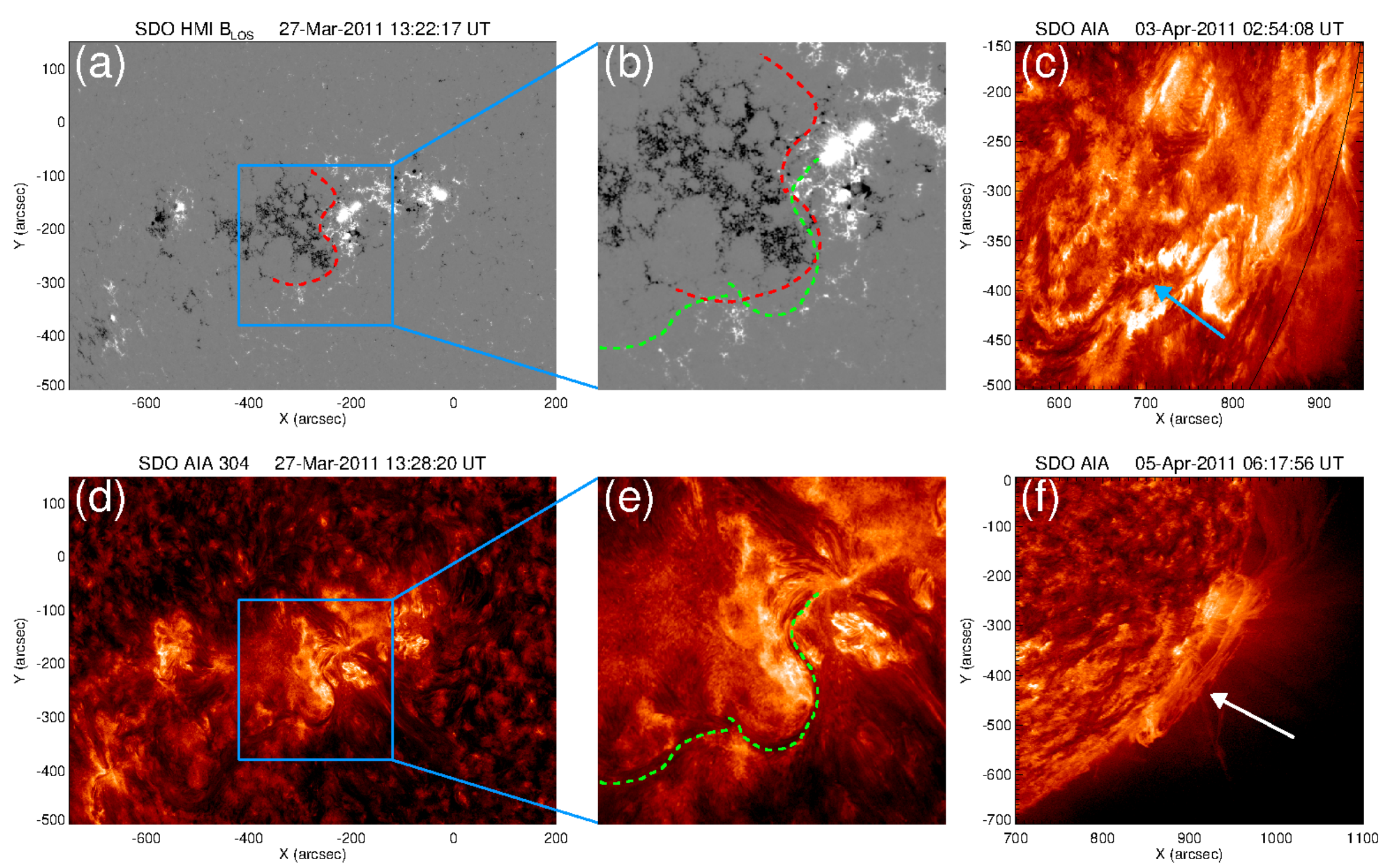}
     \caption{The HMI line-of-sight magnetic field for AR11176 plotted in gray scale within saturation values $\pm$ 500 Gauss (a). Panel (d) shows the appearance of AR11176 in AIA 304 \AA{} image within the same field-of-view as panel (a). The region bounded by the blue box in panels (a) and (d) are shown in panel (b) and (e) respectively. The red dashed line in panels (a) and (b) is the approximate  polarity inversion line drawn on top of the HMI magnetogram. The green dashed lines in panels (b) and (e) indicate the location of the filament channel as observed in AIA 304 \AA{}. Panel (c) depicts the flare ribbons indicated by the blue arrow, as observed in AIA 304 \AA{} channel during the eruption on 3 April 2011. Panel (f) shows the prominence (indicated by the white arrow) on 5 April 2011 seen in AIA 304 \AA{} channel.}
     \label{fig: Aia_AR11176}
\end{figure*}

Next, we apply the new extraction method to AR11176. The observations show that AR11176 has a bipolar-like overall structure, featuring a predominantly north-south running polarity inversion line (PIL) (see Figure \ref{fig: Aia_AR11176} (a) \& (b)). Over the course of the simulation window, there is ongoing activity close to the north-south part of the PIL, which may be tied to flux cancellation seen in the photospheric magnetic field. During the simulation window, no significant eruptions were observed, apart from localised jets forming away from the PIL as a result of flux emergence \citep{Solanki20}. However, there was an eruption that took place about two days after the end of the simulation, on 3 Apr 2011, producing clear flare ribbons along the PIL (see Figure \ref{fig: Aia_AR11176} (c)). Modelling this eruption is out of the scope of the present study due to the AR reaching too close to the limb, thereby preventing the use of the direct data-driven approach employed here. The observations suggest the presence of a filament along the PIL (see Figure \ref{fig: Aia_AR11176} (d) \& (e)). Curiously, significant parts of the filament appear to stay intact after the eruption on April 3rd, and can be seen as a prominence once it rotates to the solar limb on 5 Apr 2011 (see Figure \ref{fig: Aia_AR11176} (f)). The prominence appears to reach relatively high above the photosphere, a few tens of mega metres, suggesting a large-scale structure.

\begin{figure}
     \centering     
     \includegraphics[width=\linewidth]{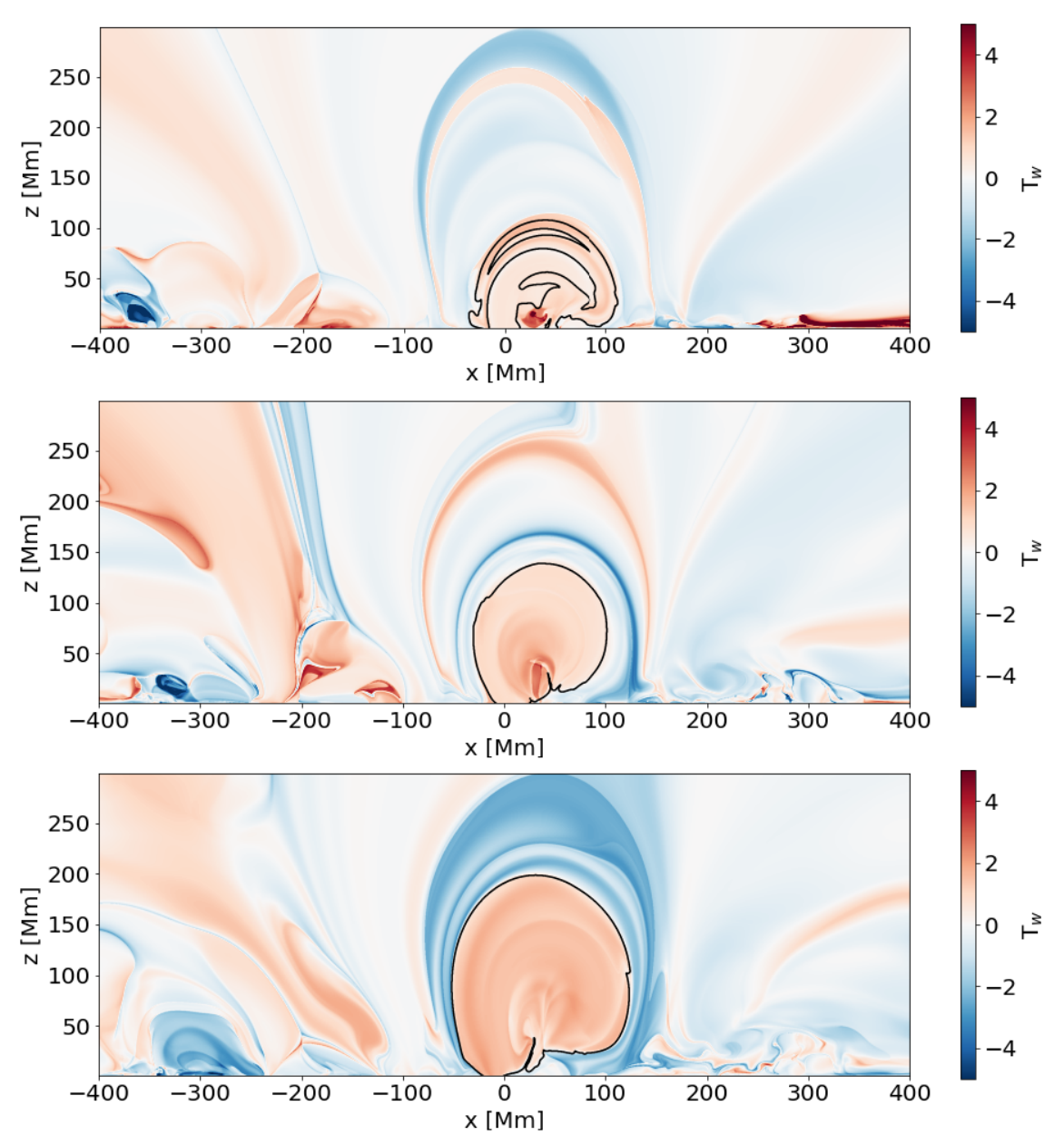}
     \caption{Same as Fig.~\ref{fig: Twist AR12473}, but for the AR11176 FR, showing simulation frames 8, 18 and 28 (from top to bottom).}
     \label{fig: Twist AR11176}
\end{figure}

The evolution of the twist maps (computed in the $y \approx -25.5$ ~Mm plane) using the AR11176 TMFM simulation is provided in Fig.~\ref{fig: Twist AR11176}. In the first panel (and associated supplementary movie) one can see the early FR formation, featuring a significantly twisted core with weakly twisted surrounding fields. As the simulation progresses, the distinction between the FR core and the surrounding field lines gradually vanishes as the high twist dissipates into the surrounding field of the same twist polarity. 
It is worth noting that the FR in AR11176 features a significantly larger cross-section upon formation than the AR12473 FR (cf., first panel of Fig.~\ref{fig: Twist AR11176} to Fig.~\ref{fig: Twist AR12473}). This also has the consequence that the apex of this FR is already at a greater height upon formation, compared to the AR12473 FR. Once the whole structure has accumulated sufficient twist such that our method finds a coherent contour (using a low threshold of $T_w \approx 0.4$ due to its less twisted nature), we identify it as FR (which happens at frame 10; cf. black contours of the top versus the middle panel in Fig.~\ref{fig: Twist AR11176}). During its evolution we observe that the structure retains its original shape rather well (middle and bottom panel), which is very different from the evolution in the simulation of AR12473. Furthermore, the FR is expanding, rather than rising, and consequently, the apex of the FR rises in its early phase significantly slower than the apex of the FR in AR12473. Note that we chose a higher domain height for the simulation of AR11176, 300 Mm instead of 200 Mm. This was done to ensure that the evolution of the FR is not influenced by interactions with the boundaries in the later stages of its evolution due to its excessive expansion.

\begin{figure}
     \centering     
     \includegraphics[width=\linewidth]{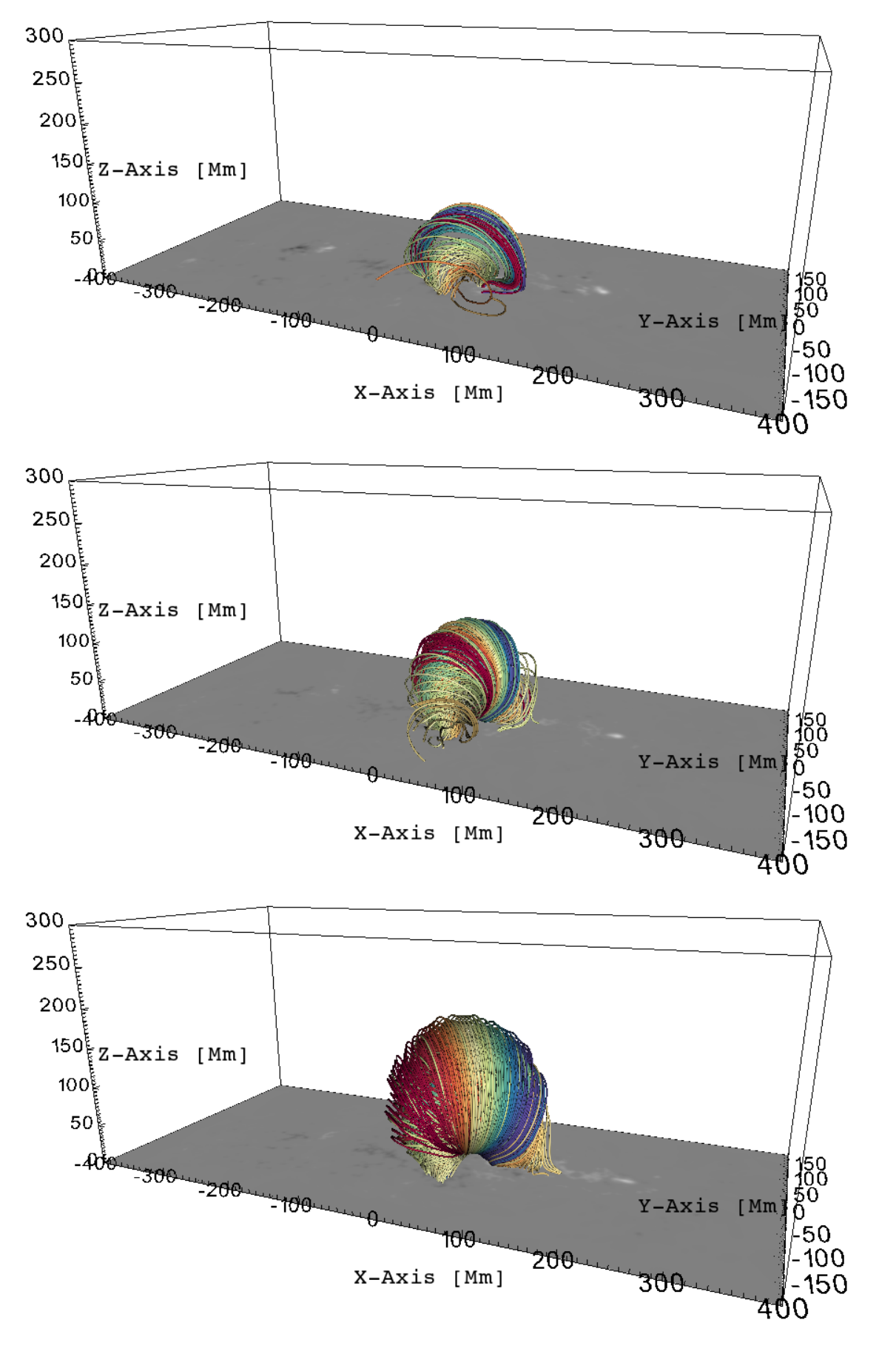}
     \caption{FR snapshots for the TMFM AR11176 simulation, using the presented extraction method. The corresponding simulation frames are 8, 18 and 28 (from top to bottom).}
     \label{fig: AR11176}
\end{figure}

The FR field lines of AR11176, extracted with the new method, are shown in Fig.~\ref{fig: AR11176}. This figure also illustrates the slow build-up of twist and its expansion, as discussed above. Figure~\ref{fig: AR11176} confirms that the FR consists (in the early to mid stages) of two parts: 1) a highly twisted core that stays approximately at the same height, and 2) a less twisted envelope that strongly expands. The height of the highly twisted FR core reaches only a few tens of megameters above the photosphere during the simulation. This elevation agrees approximately with the height of the north-south oriented prominence (Figure~\ref{fig: Aia_AR11176}). The apex of the envelope however rises to about 200 megameters, mainly due to its strong expansion. The results thus suggest that while the simulation appears to be non-eruptive, the TMFM captures the magnetic field related to the observed prominence (highly twisted core of the FR) located above the north-south running PIL.

\subsection{Circularity of Twisted Structures}
\label{Sect: Circularity}
\begin{figure}
     \centering     
     \includegraphics[width=0.8\linewidth]{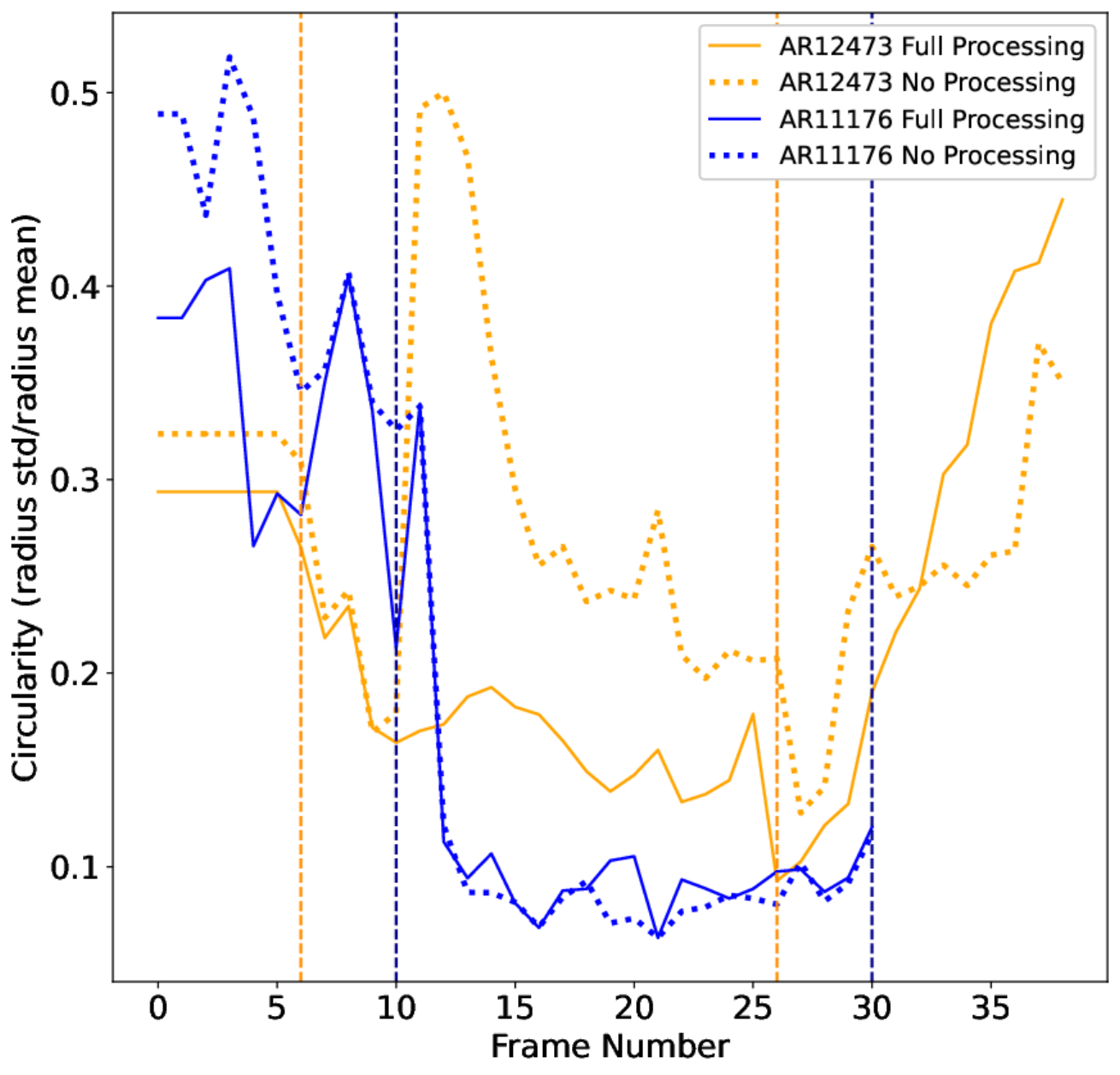}
     \caption{Circularity of the extracted cross-sectional shapes, obtained for the TMFM FRs for AR11176 (in blue) and AR12473 (in orange). The dashed vertical lines indicate the relevant (i.e., physical) time windows, which are used in Fig.~\ref{fig: Apex}. The dotted lines show the circularity of an additional FR extraction for each simulation, where no morphological openings were applied.} 
     \label{fig: circu}
\end{figure}

Next, we investigate the shape of the cross-section of the FR, determined from the twist maps using the circularity parameter (see Section~\ref{Sect: Extraction}). The obtained circularity values are displayed in Fig.~\ref{fig: circu}. The figure shows that while in the early stages the FRs have clearly a non-circular shape, they gradually become in both cases more circular as they rise/expand through the simulation domain. Note that we do exclude the very late stage evolution from the subsequent deflection analysis to mitigate tracking unphysical behaviour when the FR starts to interact with the domain boundary. Similarly, we exclude the very early stages until the FR is formed and properly detected by our method. The relevant frames are located in-between the correspondingly coloured dashed lines for both simulations.

While for the case of AR11176 the FR cross-section stabilises to a nearly circular shape, for AR12473 the FR deforms as it approaches the top of the simulation domain. We note that the sharp increase observed in Fig.~\ref{fig: circu}, at the end of the TMFM simulation for AR12473 could result from the FR deformation as it reaches the upper boundary of the simulation box and starts to exit the domain. Additional extractions, avoiding the use of morphological openings (except in one frame for each simulation, which was necessary to for a successful tracking), are shown in Fig.~\ref{fig: circu} as well. Typically, the differences in circularity are below 0.1, with the exception when a particular sub-feature of substantial size is purposefully removed. This becomes most apparent for the AR12473 simulation after frame 10 as well in the very last frames. The features that were responsible for this behaviour were long, elongated substructures in the twistmaps, that can also be seen in the middle (connection with the contour at around $y = 75$~Mm) and bottom panel (connection at approximately $z = 75$~Mm, $y = -10$~Mm) of Fig.~\ref{fig: Twist AR12473}. Cutting off these sub-features will unavoidably change the circularity of the extracted shapes, but this is necessary for correctly identifying the FR. We thus conclude that the circularity is only weakly affected by the applied morphological openings.

\subsection{Deflection Analysis}

\begin{figure}
     \centering     
     \includegraphics[width=\linewidth]{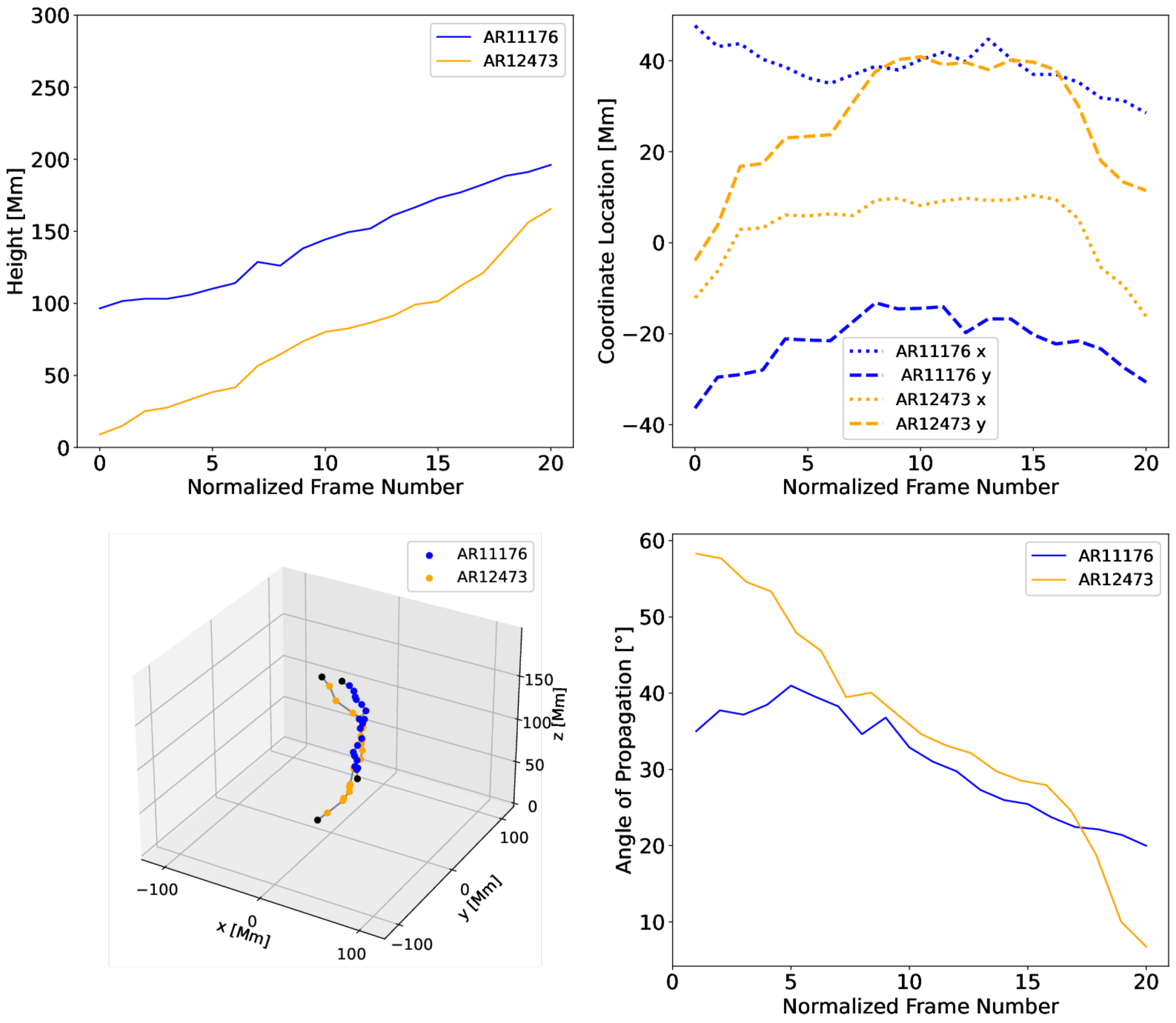}
     \caption{Trajectory of the AR12473 (orange) and AR11176 (blue) TMFM FRs. The top left panel shows the height evolution, while the top right panel shows the evolution of the horizontal components during the ascend of the FRs. The bottom left panel shows the 3D trajectory in grey, with each computed apex point marked as orange/blue dot, except the starting and end points marked in black for better visibility. The bottom right panel depicts the angle of propagation for each time step. Frame numbers are normalised to start from 0 once the FR has formed.}
     \label{fig: Apex}
\end{figure}

The results of the derived FR apex trajectories for both AR12473 and AR11176 are displayed in Fig.~\ref{fig: Apex}, with the frame numbers being normalised to start from 0, once the FRs have been formed (cf., dashed lines in Fig.~\ref{fig: circu}). From the twist maps and field line extractions shown in Figs.~\ref{fig: Twist AR12473} and \ref{fig: AR12473}, respectively, it is already clear, that the FR in AR12473 rises throughout the simulation and by the end of the simulation, a part of it has already left the simulation domain. The deflection angle analysis however shows that the trajectory of the FR features a high initial angle of propagation. The FR experiences a strong shift of propagation direction after which it stabilises for most of its further ascent. This can be seen in Fig.~\ref{fig: Apex}, both from the strong early variation in horizontal coordinates for AR12473, as well as the initially high angle of propagation, defined as in Sect.~\ref{Sect: Deflection}. This early deflection and subsequent stabilisation of the trajectory is marked by a linear evolution of the propagation angle and $x, y$ coordinates in this phase. In the later stages, the structure experiences another deflection, in approximately the inverse direction of the first one. The AR12473 (orange) graph in the bottom right panel of Fig.~\ref{fig: Apex} shows that while after the early formation stage, the angle of propagation is high (approximately $60^\circ$), the subsequent deflections counter the initial phase. In other words, the FR first propagates clearly non-radially outward but then the propagation direction changes such that the horizontal coordinates are close to its original "starting position". This results in the FR front ultimately having a very low effective deflection (ending up with an effective angle of propagation below $10^\circ$). In the bottom left panel of Fig.~\ref{fig: Apex} one can see the AR12473 FR trajectory in 3D and how the starting and ending phases practically cancel each other. 

For the FR in AR11176, on the other hand, the horizontal coordinates of the apex stay notably constant over time (in particular the x component), although the results give significant deflection angles (see blue graphs/points Fig.~\ref{fig: Apex}). Similar to AR12473, the most dramatic dynamics can be seen in the early phase. We note that we performed a linear interpolation in the spatial coordinates of the apex location at frame 17, as it was an outlier. The reason for this abrupt change of location may be the removal of some overlying  newly "formed" poloidal FR field lines (i.e, either an effect of twisting untwisted field lines and immediate reconnection thereafter or a smaller overlying twisted structure moving in and out of the 2D plane). The evolution of the FR also features a stable path of propagation in the latter phase of its evolution, which strongly departs from a purely radial direction (or in our case departs clearly from a parallel direction to the $z$-axis). This then leads to an effective angle of propagation of about $20^\circ$. The full 3D trajectory can be seen in blue in the bottom left panel of Fig.~\ref{fig: Apex}. When comparing the two simulations, one can see that while AR12473 FR rises about 150~Mm in the relevant time window, the FR in the case of AR11176 rises only 100~Mm in the same time span. 

\section{Discussion}
\label{Sect: Discussion}

\subsection{FR Evolution}
The updated FR tracking method based on the MM algorithm and the previous version from Paper~1 both extract the AR12473 FR eruption with largely similar evolutionary characteristics. The FRs start as a smaller bundle of twisted loops that continuously thicken and rise as the simulation progresses. In the later stages, two distinct structures are identified, as seen in the bottom panels of Fig.~\ref{fig: AR12473}, where the blue-colored field lines wrap around the beige and red ones in both cases. Although both tracking methods produced coherent FRs, one of the biggest differences is that the MM based extraction method captures more FR field lines and thus, appears notably thicker. Also in terms of field line connectivity, there are some differences in the early and mid phases of the simulation, particularly at the positive polarity footpoint (cf. middle panels in Fig.~\ref{fig: AR12473}). 

The height evolution is similar for the two extraction methods (cf., orange curve in the top left panel of Fig.~\ref{fig: Apex} with Fig.~3 in Paper~I), where quantitative difference appear mostly because the new method captures more FR field lines, and thus increases the apex height. In both cases the FR ascents more or less steadily through the simulation domain, until it reaches a point where the upward velocity increases. This phase is best visualised for the new algorithm in the propagation angle plots, being the orange curves in the right panels of Fig.~\ref{fig: Apex}). Due to deflection away from the intrinsic propagation direction and then back, the FR apex ends up very close to the starting direction.

The application of the new algorithm to another AR (AR11176) showed that it can capture dynamical behaviour of coronal fields in two very different types of events. In contrast to the AR12473 FR, the AR11176 FR evolves significantly slower and shows more stable behaviour, in accordance with observations. 
It forms from a twisted "core" that steadily dissipates the twist to its surrounding, forming a relatively large FR in the process. This twisted core forms at a height of about 50 Mm, while the AR12473 FR forms significantly lower in the corona, at about 10 Mm (cf.\ Figs.~\ref{fig: Twist AR12473} for AR12473 and \ref{fig: Twist AR11176} for AR11176 as well as Fig.~\ref{fig: Apex}). The AR11176 FR evolves also morphologically more steadily, as both field lines and twist maps show. Figure~\ref{fig: circu} also supports this finding, demonstrating that after the initial phase, the FR's cross-section is nearly circular (i.e., circularity values are close to zero) throughout the whole mid to late stage of the simulation. The FR consists of a set of toroidal (directed along the axis of the FR) field lines, that are partially covered by overlying field lines that connect to the same footpoint regions (implying that these overlying field lines indeed belong to the same structure), but appear as almost poloidal. 

We also found that the AR11176 FR rose more steadily (most notably in the later stages) than AR12473 FR (cf.\ Fig.~\ref{fig: Apex}), as can be seen from the lower variation of all 3 spatial coordinates and lower variability of the propagation angle. This also implies, that the FR rises significantly slower. These findings imply that the FR in the simulation of AR11176 is non-eruptive, but for a more decisive conclusion, more detailed analysis would be needed \citep[e.g., as in][]{Daei2023}. This is especially relevant because of the characteristically unrealistic slow evolution of TMFM FRs (see, e.g., \cite{Pomoell19} or Paper~I, where a temporal normalization was introduced to deal with this problem). We however note that the slower evolution of the AR11176 FR, compared to the AR12473 FR, matches well with the observations in the corresponding time windows, namely the presence of a large and non-eruptive filament/prominence that formed at the north-southward directed part of an extended PIL in the AR11176. 

\subsection{FR Extraction Method}
The new extraction scheme shows clear improvements, as discussed above and in Sect.~\ref{Sect: Extraction}. Firstly, one of the key improvements is that the updated method allows non-circular cross-sections. For the investigated events we found the flux ropes to be most circular during the mid to late rising phase, but otherwise clearly non-circular cross-sections were found. 

Secondly, although our current MM-based implementation is semi-automated (automated main extraction but case-by-case refinement of the parameters), it was not only found to be more accurate (i.e., we capture more of the relevant magnetic field lines), but also computationally more effective than the previous method. One could further speed up the MM algorithms by implementing entire sequences of erosion/dilation or opening/closing operations with small SEs instead of performing a single operation with one large SE \citep{Shih2003}. We note though that even without this possible optimisation, all computations run within seconds, using an ordinary laptop, as long as reasonable SE sizes are used. 

\subsection{Utility of MM algorithms for pre- and post-processing}
\label{Sect: Tw}

\begin{figure}
    \centering
    \includegraphics[width=\linewidth]{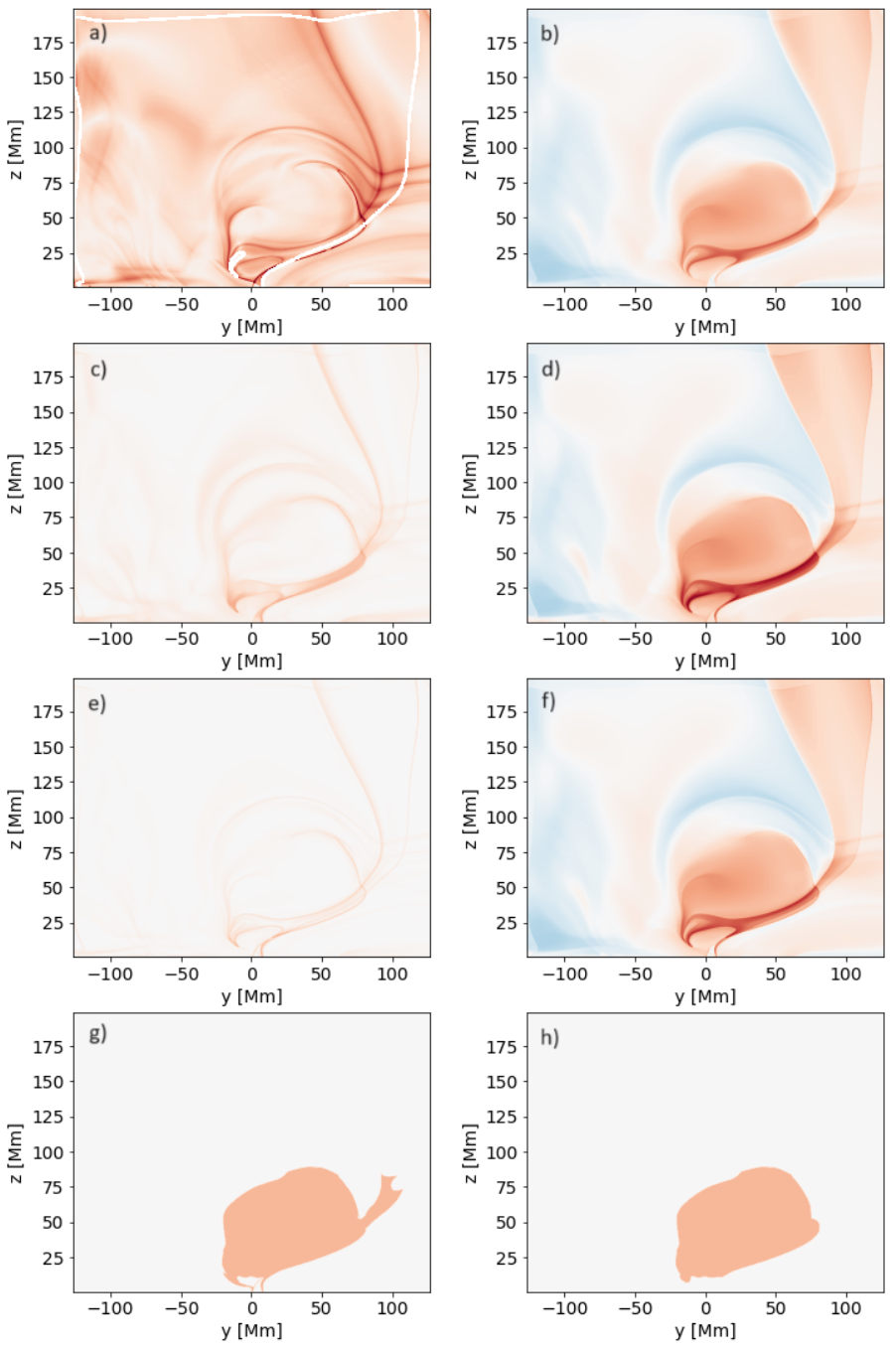}
     \caption{Comparison of different parameters and processing for frame 16 of the AR12473 TMFM simulation in the plane of $x \approx 0.7$~Mm: Panel a) shows the $\log(Q)$-map, and panel b) the twist-map. The morphological gradient using big structuring elements is shown in panel c), and its combination with the twist-map from b) is shown in panel d). Panel e) is the same as panel c), but with smaller structuring elements and panel f) is the combination of $T_w$ from panel b) with the morphological gradient in panel e). Finally, in panel g) there is the mask of high twist areas as extracted from the twist+gradient map in panel d) (using a threshold of 0.8), and panel h) shows the same mask, but after post-processing, mostly removing the twisted channel that connects to the main structure at about $y = 75\;$Mm.}
     \label{fig: TwistComp}
\end{figure}

Figure~\ref{fig: TwistComp} provides some motivation for the choice of proxy as well as insights into the extraction procedure as a whole. It is noteworthy that we do not incorporate the squashing-factor \citep[short Q-factor, see e.g.,][]{Titov2002, Demoulin1996} into our extraction scheme here. Because the extraction is based on thresholding, we require the whole FR in the extraction plane to exceed some critical value. Furthermore, it is favourable, if the FR exhibits sharp boundaries to allow for some flexibility in the choice of the threshold. This is one particular strength of $T_w$ combined with the morphological gradient: The MM gradient with different structuring elements (panel c) and e) on the left side of Fig.~\ref{fig: TwistComp}) serves as an adjustable sharpening routine to the twist maps (cf.\ original map in panel b) with the sharpened maps in panels d) and f)). The Q-factor map in panel a) appears structurally similar as $T_w$, but the outlines of highlighted features often do not fully close (see e.g., on the top right of the FR cross-section at $(y, z) = (0, 75)$ Mm). While it may still be possible to use it in combination with the twistmaps in a similar manner as the morphological gradient, it lacks the tune-ability and forces the introduction of a scaling parameter since the log(Q) maps cover different and larger ranges of values. The tune-ability with choice of different structuring elements is distinctly different from a scaling parameter as this also structurally affects the maps. Hence, we chose to use $T_w$ and the MM gradient in this work. We avoid calculating the winding number about a defined FR axis field line, as this formulation of the twist number elevates the complexity significantly and hinders a feasibly fast implementation, but note that tools for the computation of this version do exist \citep[e.g., in ][]{Price2022}{}{} and may very well be used in combination with our extraction method.

The opening algorithm is used to smooth the extracted outlines and most importantly to remove unwanted features. An example of this is the high-twist channel seen in the panels of Fig.~\ref{fig: TwistComp} that connects to the main FR body at a height of about 50~Mm and $y$-location between 50 and 100~Mm. This channel represents a set of twisted field lines that clearly do not belong to the FR. Such connections may not only lead to an incorrect extraction at this particular frame, but can cause the tracking to fail too (as in subsequent frames, the algorithm may continue to track this channel-feature instead and not the main FR in case it disconnects from it). The opening algorithm helps to disconnect this feature entirely in the affected frame. Most importantly, this processing can be applied only in the affected sub-region of the image, and thus one can keep unproblematic regions unaffected. This is illustrated by the bottom panels in Fig.~\ref{fig: TwistComp}, where one can see the original (after the extraction) on the left and the post-processed version on the right. Here, an opening with a small SE was applied to the whole image, to remove small artefacts (see e.g., near $z = 0\;$Mm), as well as an opening of bigger size to a sub-region of the image to remove the twisted "arm" of the structure. In fact, for the shown case of the AR12473 TMFM FR, the opening algorithm had to be applied for this exact reason in about half of the frames, thus it significantly contributed to improving the extraction procedure. Apart from frames, where connections like these were significant, the circularity of the extracted contours in the twistmaps were only weakly affected by the applied openings (cf., Fig.~\ref{fig: circu}). This is in strong contrast to the AR11176 FR, where significantly less post-processing was required. 

\section{Conclusion and Outlook}

In this work, we have developed and tested a new flux rope extraction and tracking method for 3D solar magnetic field simulations. The extraction is based on the twist parameter $T_w$ combined with mathematical morphology algorithms.  

We applied the new algorithm to the data-driven magnetofrictional model simulation output of the AR12473 eruption on 28 Dec 2015, i.e., the same event that was used to test the previous method  presented in \cite{Wagner2023}. To further test the new algorithm we also applied it to AR11176. We investigated both AR AR12473 and AR11176 flux ropes in terms of their appearance, motion through the modelling domain and deflection. These events presented quite different dynamics, the AR AR12473 resulting in a strong dynamics and clear flux rope eruption while the AR11176 FR was more stable in terms of its movement and we theorise it may be a confined eruption without further driving the model through the ARs main eruptive phase.

The results for both active regions are in accordance with the EUV observations. Our modelling and extraction attempts thus capture both the formation phase and early rise of the flux ropes, as well as the eruptive phase in the case of AR12473. 

While the results in Sect.~\ref{Sect: Results} illustrate that the method robustly extracts the flux ropes, fine-tuning of the parameters (probably on case-by-case basis) may still be necessary for the best extractions. Full-automation as well as exploring a broader space of flux rope proxies such as the Q-factor, winding number and magnetic flux density and possible combinations of the aforementioned quantities with each other as well as with mathematical morphology algorithms are avenues for further improvements. 

\begin{acknowledgements}
This work is part of the SWATNet project funded by the European Union’s Horizon 2020 research and innovation programme under the Marie Skłodowska-Curie grant agreement No 955620. The work is supported by Academy of Finland Centre of Excellence \mbox{FORESAIL} (grant 336807). EK and AK acknowledges European Research Council under the European Union’s Horizon 2020 research and innovation programme, grant 724391 (SolMAG). RS acknowledges support from the project EFESIS under the Academy of Finland Grant 350015. TB and RG acknowledge the support by Funda\c{c}\~{a}o para a Ci\^{e}ncia e a Tecnologia (FCT) through the research grants UIDB/04434/2020 and UIDP/04434/2020. 
R.E.\ is grateful to Science and Technology Facilities Council (STFC, grant No. ST/M000826/1) UK and acknowledges NKFIH (OTKA, grant No. K142987) Hungary for enabling this research. JP acknowledges Academy of Finland project SWATCH (343581).
We acknowledge the use of the data from the Solar Dynamics Observatory (SDO). SDO is the first mission of NASA's Living With a Star (LWS) program.
The SDO/AIA and SDO/HMI data are publicly available from NASA's SDO website https://sdo.gsfc.nasa.gov/data/).
O.O. acknowledges support by the FCT – Funda\c{c}\~{a}o para a Ci\^{e}ncia e a Tecnologia, I.P., undeProjects Nos. UIDB/04564/2020, UIDP/04564/2020 and CERN/FIS-COM/0029/2017. AK’s research was supported by an appointment to the NASA Postdoctoral Program at the the NASA Goddard Space Flight Center (GSFC). AK acknowledges the University of Helsinki Three-year Grant.
S.P.\ acknowledges support form the projects
C14/19/089  (C1 project Internal Funds KU Leuven), G.0B58.23N and G.0025.23N (WEAVE)   (FWO-Vlaanderen), 4000134474 (SIDC Data Exploitation, ESA Prodex-12), and Belspo project B2/191/P1/SWiM. We also want to thank the referee for their feedback, which lead to improving the manuscript.
\end{acknowledgements}

\bibliographystyle{aa}
\bibliography{Bibfile}

\end{document}